\documentclass[graybox]{svmult}

\usepackage{mathptmx}       
\usepackage{helvet}         
\usepackage{courier}        
\usepackage{type1cm}        
\usepackage{makeidx}         
\usepackage{graphicx}        
\usepackage{epsf}
\usepackage{multicol}        
\usepackage[bottom]{footmisc}

\makeindex             

\begin{document}

\title*{Electronic charge and orbital reconstruction at cuprate-titanate interfaces}
\author{Natalia Pavlenko and Thilo Kopp}
\institute{Natalia Pavlenko \at Center for Electronic Correlations and Magnetism (EKM), Universit\"at Augsburg, 86135 Augsburg, 
Germany; 
\at Institute for Consensed Matter Physics, 79011 Lviv, Ukraine, \email{pavlenko@mailaps.org}
\and Thilo Kopp \at Center for Electronic Correlations and Magnetism (EKM), 
Universit\"at Augsburg, 86135 Augsburg, Germany}
\maketitle

\abstract{In complex transition metal oxide heterostructures of physically dissimilar perovskite 
compounds, interface phenomena can lead to novel physical properties not observed
in either of their constituents. This remarkable feature opens new
prospects for technological applications in oxide electronic devices based
on nm-thin oxide films. Here we report on a significant electronic charge
and orbital reconstruction at interfaces between YBa$_2$Cu$_3$O$_{6}$ and SrTiO$_3$
studied using local spin density approximation (LSDA) with intra-atomic Coulomb repulsion
(LSDA+$U$). We show that the interface polarity results in the metallicity of cuprate-titanate
superlattices with the hole carriers concentrated predominantly in the CuO$_2$ and BaO layers and in
the first interface TiO$_2$ and SrO planes. We also find that the interface structural relaxation causes a 
strong change of orbital occupation of Cu $3d$ orbitals in the CuO$_2$ layers. The
concomitant change of Cu valency from $+2$ to $+3$ is related to the partial occupation 
of the Cu $3d_{3z^2-r^2}$ orbitals at the interface with SrO planes terminating SrTiO$_3$. 
Interface-induced predoping and orbital reconstruction in CuO$_2$ layers are key mechanisms 
which control the superconducting properties 
of field-effect devices developed on the basis of cuprate-titanate heterostructures.
}

\abstract*{In complex transition metal oxide heterostructures of physically dissimilar perovskite
compounds, interface phenomena can lead to novel physical properties not observed
in either of their constituents. This remarkable feature opens new
prospects for technological applications in oxide electronic devices based
on nm-thin oxide films. Here we report on a significant electronic charge
and orbital reconstruction at interfaces between YBa$_2$Cu$_3$O$_{6}$ and SrTiO$_3$
studied using local spin density approximation (LSDA) with intra-atomic Coulomb repulsion
(LSDA+$U$). We show that the interface polarity results in the metallicity of cuprate-titanate
superlattices with the hole carriers concentrated predominantly in the CuO$_2$ and BaO layers and in
the first interface TiO$_2$ and SrO planes. We also find that the interface structural relaxation causes a
strong change of orbital occupation of Cu $3d$ orbitals in the CuO$_2$ layers. The
concomitant change of Cu valency from $+2$ to $+3$ is related to the partial occupation
of the Cu $3d_{3z^2-r^2}$ orbitals at the interface with SrO planes terminating SrTiO$_3$.
Interface-induced predoping and orbital reconstruction in CuO$_2$ layers are key mechanisms
which control the superconducting properties
of field-effect devices developed on the basis of cuprate-titanate heterostructures.
}

\section{Introduction}\label{intro}

It is well known that the rich properties of transition metal oxides like ferroelecticity, magnetism or
superconductivity are closely associated with the physics of $d$-orbitals \cite{cox}. In heterostructures of physically
different transition metal oxides, the local structural deformations and uncompensated charge at the interfaces
result in properties well beyond a simple combination of the characteristics of their constituents \cite{ahn2,kiguchi,thiel}. 
A prominent example is the titanate 
superlattice of insulating perovskites SrTiO$_3$ and LaTiO$_3$ where the metallic conductivity
is caused by the mixed valence ($+3/+4$) of Ti and by electronic charge reconstruction at polar interfaces \cite{ohtomo}.
The new electron states appearing in such heterostructures can be tuned by external electromagnetic fields
which opens new prospects for engineered oxide electronic devices based on nm-thin transition metal oxide films.

In heterostructures with high-$T_c$ cuprate films, the external electrostatic fields tune the 
surface/interface charge density and in this way lead to transitions between 
conducting (or superconducting) and insulating states. Such heterostructures consist of YBa$_2$Cu$_3$O$_{7-\delta}$-films 
grown on SrTiO$_3$ layers and are of central importance in superconducting field-effect devices \cite{mannhart,ahn,logvenov,cassinesse}. 
In the existing literature,
the shift of $T_c$ achieved with increasing electrostatic field is often understood in terms of
an electrostatic doping concept without detailed consideration of the microstructure of the interfaces
between YBa$_2$Cu$_3$O$_{7-\delta}$-film and SrTiO$_3$ \cite{ahn,logvenov}. Several experimental facts, however, indicate an interface-related 
modification of the electronic states in the cuprate/perovskite oxide heterostructures. For example,
recent studies on underdoped cuprate films produced a $T_c$-shift of about
5--15~K, whereas in the overdoped films no shifts were observed, a fact, which 
cannot be explained satisfactorily by field-doping \cite{logvenov}. Despite 
the important role of interfaces in the 
physical properties of such oxide heterostructures, no direct experimental studies 
of the electronic properties of interfaces between copper and titanium oxides are currently 
available. This is partially related to extreme experimental
difficulties in distinguishing the direct interface- and bulk-contributions in the experimental data \cite{abbamonte}. 
On the other hand, due to the high complexity of the cuprate and titanate structures, 
extensive  theoretical studies of the cuprate-titanate interfaces \cite{wehrli,pavlenko2,pavlenko_kopp,pavlenko3,koerting,pavlenko4}
are a challenging task which requires extremely demanding computational
resources \cite{pavlenko,noguera}. 

The goal of the current project is the development of interface models 
for microstructures of strongly correlated cuprates and titanates. The 
analyses of novel electronic states are based on the density functional theory within LSDA and LSDA+$U$ implementations.
Due to the insufficient structural information about
atomic arrangements at cuprate-titanate interfaces,  the theoretical investigations 
include the theoretical modeling of heterostructures with different structurally compatible surface terminations of SrTiO$_3$ and YBa$_2$Cu$_3$O$_{6}$
and relaxation of interface atomic 
structures.  In this context, the main amount of the computer time is used for high-performance theoretical engineering of multilayer 
oxide supercells and for structural optimization. 

In our extensive analyses, we find a metallic state in cuprate-titanate interfaces which is characterized by 
significant hole predoping of cuprate films \cite{pavlenko}. In the field-effect experiments, such an interface-caused predoping 
occurs even before the electrostatic injection of charge and is of essential importance for functional properties
of superconducting field-effect devices. Furthermore, out results not only allow to obtain a clear answer
to the question about the electronic charging of the cuprate-titanate heterostructures, but also to make important
predictions how to increase the performance of field-effect devices.
Our studies directly show that despite the extreme computational resources required for optimization procedures, 
the interface structural relaxation is of crucial importance
for understanding the physics of transition metal oxide heterostructures. In the copper oxide layers
of cuprate-titanate superlattices, we find that the interface-caused local structural 
deformations strongly affect electronic occupancies of $d$-orbitals of Cu and can lead to 
a change of its electronic valence state. 
Due to the central role of $d$-orbitals in the magnetic
and superconducting properties of transition metal oxides, the obtained interface-induced orbital reconstruction
should have dramatic consequences for the functionality of field-effect devices based on cuprate-titanate superlattices.      

\section{Description of the methods}

Our extensive studies of oxide surfaces are based on the density functional theory within 
the local spin density approximation (LSDA) 
with intra-atomic Coulomb repulsion (LSDA+$U$) \cite{anisimov}. The calculations of the electronic densities of states,
structural relaxation in terms of minimization of total energy and forces have been performed using 
the linearized augmented plane wave method (LAPW) implemented in the WIEN2k package \cite{wien2k}. The most numerically
intensive parts of the package code are executed in parallel, namely full potential calculations, iterative solution of eigenvalues
and eigenvector problems of 
the generated Hamiltonian matrix and calculations of electronic charge density expansions. Parallelization
is achieved on the $\vec{k}$--point level by distributing subsets of the k-mesh to different processors and is based on parallelization libraries
including MPI, Scalapack, and BLACS. 

The small amount of the computational resources was spent for the development of interface-related extensions to the WIEN2K-code
which are aimed at specific studies of the local electronic
structure and the charge density profiles across the interfaces in heterostructures. For instance, to obtain the profiles of the
hole density across the polar interfaces, we had to generate the charge densities in the energy window between the Fermi
level and the top of the valence bond and to perform a subsequent planar integration of these profiles. In this way, not only the
charge distribution, but also the effective thickness of the 
interface, where the charge density deviates from its usual bulk-type behavior, can be 
estimated.

The main part of the
resources was absorbed by computationally demanding structural optimization and LDA+$U$-calculations of the optimized superlattices.
Specifically, for superlattices of YBa$_2$Cu$_3$O$_6$/SrTiO$_3$ and CuO$_2$/SrTiO$_3$ containing from 18 to 22 atoms in a supercell,
a single run of LDA+$U$ included from 50 to 80 iteration steps until the final convergence with respect to the electronic charge and total energy
could be reached. With  
15--20 processors used for the parallel $k$-points calculations, each converged LDA+$U$ run usually required 
up to 25,000 CPU hours. In addition, in each superlattice, the relaxation of the interface distances 
could be achieved by performing up to 10 single LSDA runs which required about 10,000 CPU hours.  Due to a wide range of superlattices 
analyzed in our project (YBCO/STO-sandwiches, CuO$_2$/STO-slubs, Sr$_2$CuO$_2$Cl$_2$/STO superlattices), such computationally demanding
calculations have used about 150,000 CPU hours in the first 6 months after beginning the project. 

The further stages of calculations of interfaces require a full optimization of all local atomic coordinates in the superlattices
and the increase of the number of elementary unit cells of 
the constituent compounds in each supercell. Consequently, the computational time
required for a structural optimization 
drastically increased up to 25,000 CPU hours for a single superlattice, whereas each LDA+$U$ 
single run requires about 
35--40,000 CPU hours for a supercell with 33 different atoms. Without the exceptional computational resources provided by HLRB, 
the performance of such demanding calculations would not be a feasible task 
and could not have been achieved in a realistic time period.  

\section{Electronic properties of cuprate-titanate superlattices}\label{ct}

The recent X-ray studies of the interface arrangement
\cite{bals,abbamonte} give clear indications of incompletely grown unit cells
of the cuprate film on the SrTiO$_3$ substrate. Whereas the bulk unit cell
of YBa$_2$Cu$_3$O$_{7-\delta}$ is terminated by CuO chains, 
responsible for the charge doping of cuprates, the CuO chains near 
the interface are missing which makes the corresponding unit cell incomplete.
This structural rearrangement should strongly affect the superconducting properties of cuprate thin films. 
Therefore, we focus in our studies on
the consequence of the effective `substitution' of electrostatically charged CuO-chains 
with electrostatically neutral (001) layers of SrTiO$_3$. As the superlattices, 
which result from such a structural rearrangement, will be polar, the 
overall charge neutrality requires extra hole carriers 
to be distributed near the cuprate-titanate interface. Our results
clearly show that in YBa$_2$Cu$_3$O$_{6}$-SrTiO$_3$ heterostructures
such interface induced hole predoping may drastically change the interfacial electronic 
properties.

In YBa$_2$Cu$_3$O$_{6}$-SrTiO$_3$ heterostructures, the chemical bonding at the 
(001) interface with the cuprate film will be determined by the first termination plane
of SrTiO$_3$, which can be either TiO$_2$ or SrO. 
Below we demonstrate the consequences of interface electronic reconstruction
for these two basic cases of SrTiO$_3$ termination.

\subsection{Case study: TiO$_2$ termination at interface}

\begin{figure}
\epsfxsize=6.5cm \centerline{\epsffile{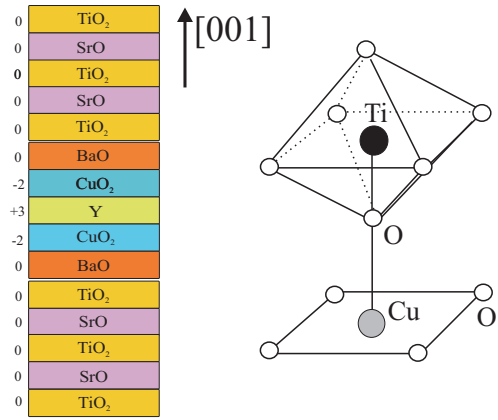}} \caption{Scheme of a
YBaCuO/SrTiO$_3$-sandwich where the polar interfaces appear due to the
incomplete Ba$_2$Cu$_2$O$_6$-unit cell with an interface structural   
configuration shown in detail on the right.} \label{fig1}
\end{figure}

In the case when the substrate of SrTiO$_3$ is  
terminated by a TiO$_2$-layer, the determined interface arrangement is
typically a stack of
$\ldots$/TiO$_2$/BaO/CuO$_2$/Y/CuO$_2$/$\ldots$ layers. Such
structural stacks suggest an interface chemical bonding Ti-O-Cu with the oxygen
of the BaO-layers shared between the CuO$_2$ and TiO$_2$-planes. From the  
electrostatical point of view, the initial `bulk-type' electronic charging of
the constituent layers indicated in the left panel of Fig.~\ref{fig1} would
result in 1 extra hole which is needed in order to compensate the polarity. 
In the sandwich-type supercell shown in Fig.~\ref{fig1}, this compensation 
leads to a doping of upper and lower symmetric 
blocks $\ldots$/SrO/TiO$_2$/BaO/CuO$_2$ by 0.5 holes. 

For the YBaCuO/SrTiO sandwich-type supercells, we performed calculations of
the electronic structure using the SIC variant of the LSDA+$U$ method 
on a $9\times 9\times 1$ $\vec{k}$--point grid with $U=8$~eV and $J=0.8$~eV
on the Cu $3d$ orbitals. The lattice constants of the supercell 
were fixed to the lattice constant $a=b=3.898$~\AA~of cubic SrTiO$_3$. 
The interface distance  $\Delta$ between the nearest BaO and TiO$_2$ planes
was optimized by minimization of the total energy which leads
to the value $\Delta=1.85$~\AA~between the apical oxygens of BaO and
the TiO$_2$ plane.

\begin{figure}[t]
\epsfxsize=7.5cm \centerline{\epsffile{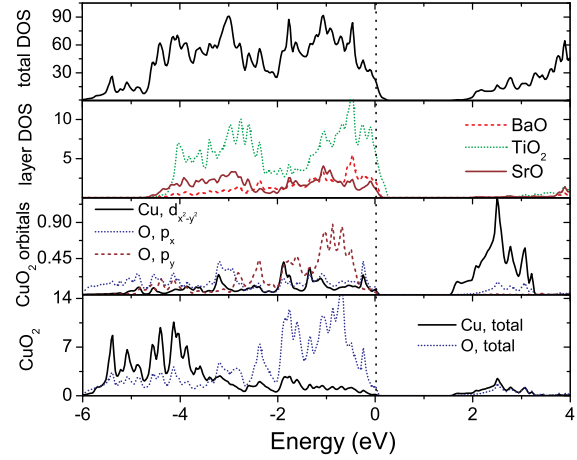}} \caption{Density of states of
the SrTiO$_3$/YBa$_2$Cu$_2$O$_6$/SrTiO$_3$-sandwich calculated within the LSDA+$U$
approach with $U=8$~eV and $J=0.8$~eV for the electrons in Cu $3d$ orbitals.
The zero of energy is at the Fermi level.} \label{fig2}
\end{figure}

From the calculated density of states, shown in Fig.~\ref{fig2}, we can
identify the metallic state with hole carriers 
originating from the oxygen $p$ states. Like in the bulk YBa$_2$Cu$_3$O$_6$, the
Cu $d_{x^2-y^2}$ states are separated by a gap of $1.34$~eV from
the oxygen $2p$ states.
It should also be noted that for this type of interface
arrangement, the orbitals Cu $d_{3z^2-r^2}$ and $t_{2g}$ are occupied and
remain below the Fermi level. Fig.~\ref{fig2} demonstrates that the significant amount
of hole charge is located in the interfacial TiO$_2$, BaO and more distant SrO layers.
Also, a substantial amount of hole charge is distributed over the CuO$_2$ planes.

\begin{figure}
\epsfxsize=8.5cm \centerline{\epsffile{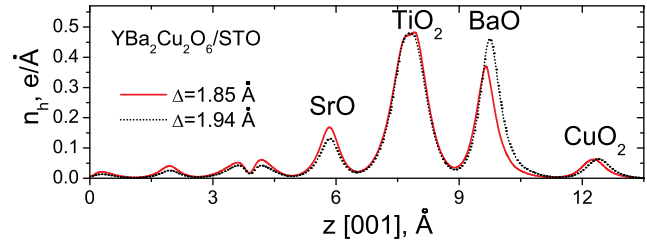}} \caption{Hole density
distribution near the YBa$_2$Cu$_2$O$_6$/SrTiO$_3$-interface for optimized
($\Delta=1.85$~\AA) and unrelaxed ($\Delta=1.94$~\AA) cases. The position $z=0$
is at the bottom TiO$_2$-plane of
the SrTiO$_3$/YBa$_2$Cu$_2$O$_6$/SrTiO$_3$-sandwich.} \label{fig3}
\end{figure}

More detailed information about the distribution of 
the charge-compensating hole density is presented in Fig.~\ref{fig3}.
Through an integration of the hole carrier density  
along the interfacial $z$-direction we obtain that the major part of hole carriers 
is distributed along the BaO (25$\%$) and the first TiO$_2$ ($48\%$) and SrO ($12\%$) layers.
Furthermore, about $5 \%$ of the total interface hole charge $0.5$~e is located in CuO$_2$ planes.
This suggests a finite metallic conductivity in the titanate, BaO, and copper oxide planes.
It is noteworthy that the structural relaxation for this type of interface leads to a reduction
of the hole density in the BaO plane and to its slight redistribution within the interface SrO and 
more distance planes, whereas the hole density and orbital occupation in the CuO$_2$ planes  
remains almost unaffected.

Our results for the hole charge in the copper oxide planes clearly show that, apart from chemical
doping, the interface polarity is another decisive factor that modulates the doping level
in the cuprate films. In superconducting field-effect devices, operated by electrostatic charging, 
such initial interface-induced predoping 
levels may have dramatic consequences for their
performance. From the point of view of superconductivity, the most important feature is the shift 
of $T_c$ with electrostatic doping which will be directly affected by the obtained predoping ($x=0.025$)
in the copper oxide film. Moreover, in other structural configurations
at YBCO/SrTiO interfaces, much higher hole predoping levels and dramatic changes of 
electronic occupancies of Cu $3d$ orbitals in CuO$_2$ planes can appear as a result of electronic and orbital
interface reconstruction.    

\subsection{Case study: SrO termination at interface}

\begin{figure}[t]
\epsfxsize=7.0cm \centerline{\epsffile{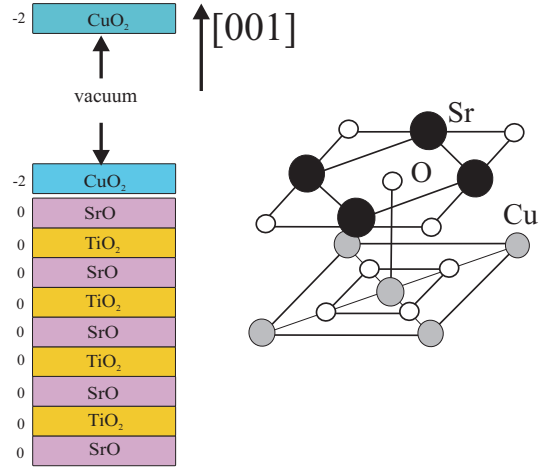}} \caption{Scheme of a polar
CuO$_2$/SrTiO$_3$-superlattice where a STO-layer is terminated by a SrO-plane.
The right side shows a structural configuration which appears at the
interface.} \label{fig4}
\end{figure}

To analyze 
the structurally different SrO termination of the substrate SrTiO$_3$, we consider
a model case in which a copper oxide plane is directly deposited on SrTiO$_3$ (Fig.~\ref{fig4}).
The direct deposition of the single Cu$^{2+}$O$^{4-}$ plane on the electrostatically neutral
titanate layer would require two extra holes to maintain the 
overall charge neutrality.
To achieve such an extremely high doping level, interface electronic reconstruction is inevitably
required. Apart from the electronic mechanism, other forms of interface reconstruction 
like oxygen vacancies or cation intermixing could modify the chemical composition. However, it is
still instructive to enforce atomically flat and stoichiometric surfaces in order to study 
comprehensively different mechanisms of the electronic reconstruction.

In our work, in order to focus on the effect of electronic reconstruction, we have introduced a 
decoupling vacuum layer of $13$~\AA~thickness between the CuO$_2$ surfaces in the slub geometry 
shown in Fig.~\ref{fig4}. We have also performed an optimization of the superlattice structure
where a relaxed interface distance $\Delta=1.83$~\AA~between the CuO$_2$ and SrO corresponds
to a minimum of the total energy. 

\begin{figure}
\epsfxsize=8.5cm \centerline{\epsffile{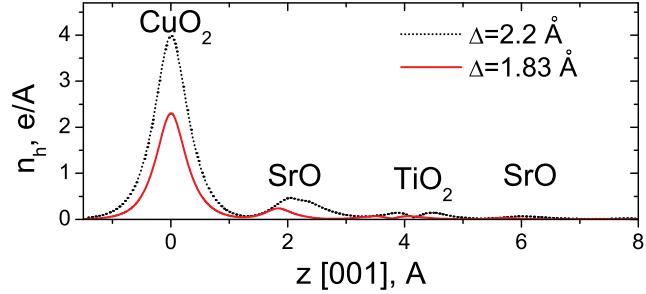}} \caption{Hole density
distribution in the interface planes of the CuO$_2$/SrTiO$_3$ superlattice.
Here $z=0$ corresponds to the location of the lower CuO$_2$-plane.}
\label{fig5}
\end{figure}

In contrast to the case of TiO$_2$ termination, 
the effect of the structural relaxation on the distribution of the interface-predoped hole density is crucial 
for the CuO$_2$/SrO interface (Fig.~\ref{fig5}).
To demonstrate this fact, we show in Fig.~\ref{fig5} the distribution of hole carrier density for the optimized
($\Delta=1.83$~\AA) and unrelaxed ($\Delta=2.2$~\AA) interfaces. The density plots across the interface 
are obtained by the integration  
of generated charge density in the energy window between the Fermi level and the top of
the O $2p$ bands.
As can be seen in Fig.~\ref{fig5}, the relaxation of the interface distance $\Delta$ leads to a dramatic
decrease of the hole density $n_h$ in the CuO$_2$ layers. The integration of $n_h$ along the interfacial $z$
direction shows that the total predoped hole charge $N_h$ accumulated predominantly in the $2p$ orbitals of O and
hybridized with $3d$ Cu orbitals, amounts exactly 
to 2 holes in the unrelaxed ($\Delta=2.2$~\AA) supercell.
In contrast to this, in the relaxed structure the density integration gives $N_h=1$ hole for the total 
hole charge. Such 
a strong difference between $N_h$ in relaxed and unrelaxed cases appears due to the additional 
orbital mechanism of the polarity compensation. It becomes important due to the interface structural 
relaxation.  

\begin{figure}
\epsfxsize=11.0cm \centerline{\epsffile{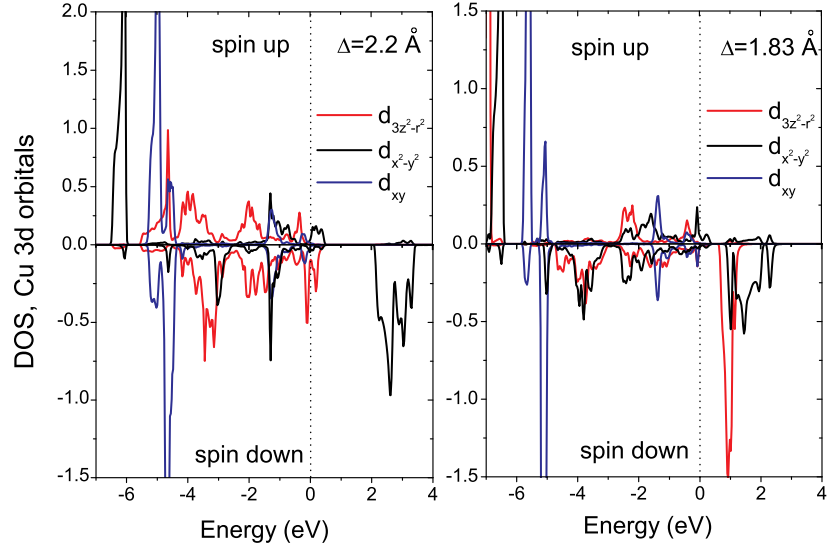}} \caption{Projected Cu $3d$
orbital density of states at the interface between CuO$_2$ and SrTiO$_3$
terminated by SrO (LSDA+$U$ studies). The zero of energy is at the Fermi level.
The left and right panels correspond to the case of unrelaxed
($\Delta=2.2$~\AA) and optimized ($\Delta=1.83$~\AA) interfacial distances.} \label{fig6}
\end{figure}

The new mechanism of the orbital reconstruction becomes evident when we compare the partial densities
of states of $3d$ orbitals calculated for optimized and unrelaxed supercells demonstrated in Fig.~\ref{fig6}. 
The figure shows that the $3d_{x^2-y^2}$ and $3d_{xy}$ orbitals of Cu are affected only slightly
by the decrease of
the [CuO$_2$-SrO] distance $\Delta$. Specifically, the orbitals $3d_{xy}$ are fully occupied and located
well below the Fermi level independently of $\Delta$. Furthermore, a significant amount of holes is 
found in the states $3d_{x^2-y^2}$ which are split due 
to the charge transfer gap of about $1.5$~eV.
The charge transfer gap becomes narrower with the decrease of $\Delta$. The most striking effect, 
which appears for the relaxed $\Delta=1.83$~\AA, is the 
reconstruction of Cu $3d_{3z^2-r^2}$ orbitals. While for 
$\Delta=2.2$~\AA\ both Cu $3d_{3z^2-r^2}$ orbitals are located
about $4-5$~eV below the Fermi level, the decrease of $\Delta$ leads to their splitting. This splitting results in the 
``displacement'' of one $d_{3z^2-r^2}$ orbital (spin down case in Fig.~\ref{fig6}) 
to about $1$~eV above the Fermi level
which implies that this orbital becomes empty. Such a strong change of the occupation of $d_{3z^2-r^2}$ corresponds to 
an increase of the valency of Cu from $+2$ to $+3$.
It occurs due to the change of the 
Coulomb potential for the decreasing [CuO$_2$-SrO] distance and the corresponding deformation
of interfacial CuO$_6$ octahedra. As a result of the orbital reconstruction, the local electron
configurations of Cu atoms in the antiferromagnetically ordered interface CuO$_2$ planes can be
represented by an alterating sequence of $\{ 3d^{5\uparrow 3\downarrow}; 3d^{5\downarrow 3\uparrow} \}$ local
spin states. Such an arrangement leads to an enhancement of the local magnetic moment on Cu up 
to $33\%$ which is also consistent with the corresponding Hund rules of electronic orbital arrangement.

Consequently, at the CuO$_2$-SrO interfaces, two basic reconstruction mechanisms are 
involved in order to compensate the interface polarity: (i) electronic charge compensation which leads to the hole
charge predoping of total amount $N_h=1$ with holes located predominantly in the 
O $2p$ orbitals
hybridized with Cu $3d_{x^2-y^2}$ in the CuO$_2$ planes; (ii) orbital reconstruction associated with 
changes of the occupancy of Cu $3d_{3z^2-r^2}$ orbitals so that exactly one of these orbitals becomes unoccupied by electrons.    
In contrast to the unrelaxed interfaces where only the first charge-predoping mechanism is responsible for
the electrostatic neutrality, the relaxation of the neutral supercell leads to a combination of two 
(charge and orbital) mechanisms which become equally important in order to achieve the stability of the system. 

\begin{figure}
\epsfxsize=8.5cm \centerline{\epsffile{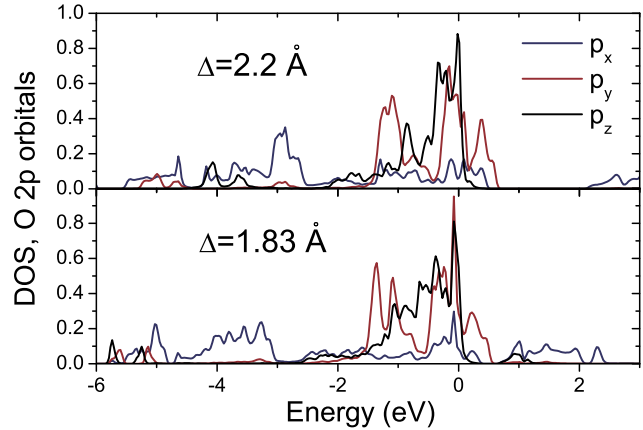}} \caption{Projected O $2d$
orbital density of states in the CuO$_2$ planes at the interface with SrTiO$_3$
terminated by SrO (LSDA+$U$ studies). The zero of energy is at the Fermi level.
The top and bottom panels correspond to the case of unrelaxed
($\Delta=2.2$~\AA) and optimized ($\Delta=1.83$~\AA) interfacial distances.
}
\label{fig7}
\end{figure}

\begin{figure}
\epsfxsize=8.5cm \centerline{\epsffile{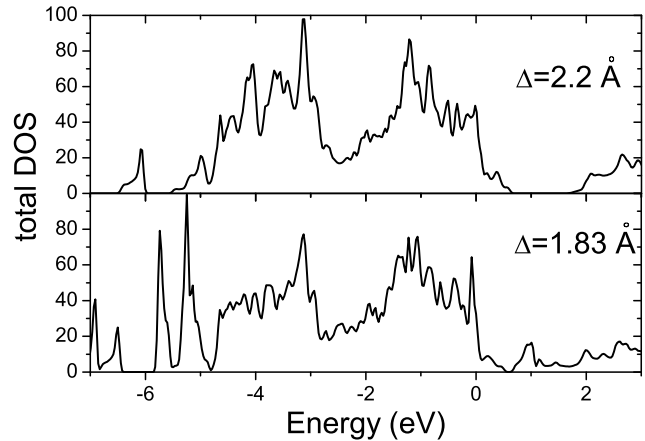}} \caption{Total density of
states of the superlattice with CuO$_2$ deposited on SrTiO$_3$, terminated by
SrO (LSDA+$U$ studies). The zero of energy is at the Fermi level. The top and
bottom panels correspond to the case of unrelaxed ($\Delta=2.2$~\AA) and
optimized ($\Delta=1.83$~\AA) interfacial distances.}
\label{fig8}
\end{figure}

The new reconstruction mechanism also involves a redistribution of 
the partial density
of states for the O $2p$ orbitals in the CuO$_2$ planes as shown in Fig.~\ref{fig7}. In the unrelaxed heterostructure
the Fermi level is located at a distance of about $0.7$~eV below the top of 
the O $2p$ bands implying high hole densities.
The interface relaxation 
leads to a shift of the Fermi level closer to the top of the $2p$ bands which leads to the decrease of the hole charge.
It is remarkable that such a modification of the structure near the Fermi level is also accompanied by a significant reduction of
the energy gap between O $2p$ and Cu $3d$ orbitals, 
an effect which is observed in the total DOS, displayed in Fig.~\ref{fig8}.
Such a reduction of the gap is caused by the location of the empty Cu $3d_{x^2-y^2}$ band at a distance less then $1$~eV above 
the Fermi level which is a clear manifestation of the orbital reconstruction mechanism in the structurally
relaxed heterostructure.   

It should be noted that up to now, the growth of high quality YBaCuO films on the substrates
of structurally compatible transition metal oxides 
remains a challenging task due to their roughness caused partially by ionic compensation of
interface polarities. The existing difficulties with assembling heterostructures as well as
with probing the interfacial physics directly make the theoretical calculations a powerful
alternative tool for studies of oxide interfaces. In fact, the 
importance of the presented results is 
supported by recent resonant X-ray spectroscopic studies of 
interfaces between YBa$_2$Cu$_3$O$_7$ and La$_{2/3}$Ca$_{1/3}$MnO$_3$ \cite{chakhalian}
for which evidence of orbital reconstruction with partial occupation of 
Cu $3d_{3z^2-r^2}$ orbitals has been provided. 
Our theoretical studies identify a possible mechanism
of such orbital reconstructions. In this context, the presented HLRB project contributes new and fascinating results 
in the rapidly developing field of the physics of transition metal oxide surfaces/interfaces.  

\bigskip
\noindent {\it Acknowledgements.}
This work was supported by the HLRB project h1181 of the Leibniz-Rechenzentrum M\"unchen and the DFG through SFB 484.

\end{document}